# Improved Contacts to MoS$_2$ Transistors by Ultra-High Vacuum Metal Deposition

Chris D. English[1], Gautam Shine[1], Vincent E. Dorgan[2], Krishna C. Saraswat[1], Eric Pop[1,3]

[1]*Electrical Engineering, Stanford University, Stanford, CA, 94305, USA*
[2]*Electrical & Computer Eng., Univ. of Illinois at Urbana-Champaign, Urbana, IL 61801, USA*
[3]*Precourt Institute of Energy, Stanford University, Stanford, CA, 94305, USA*

Contact: epop@stanford.edu

**ABSTRACT:** The scaling of transistors to sub-10 nm dimensions is strongly limited by their contact resistance ($R_C$). Here we present a systematic study of scaling MoS$_2$ devices and contacts with varying electrode metals and controlled deposition conditions, over a wide range of temperatures (80 to 500 K), carrier densities ($10^{12}$ to $10^{13}$ cm$^{-2}$), and contact dimensions (20 to 500 nm). We uncover that Au deposited in ultra-high vacuum (~$10^{-9}$ Torr) yields three times lower $R_C$ than under normal conditions, reaching 740 Ω·μm and specific contact resistivity $3\times10^{-7}$ Ω·cm$^2$, stable for over four months. Modeling reveals separate $R_C$ contributions from the Schottky barrier and the series access resistance, providing key insights on how to further improve scaling of MoS$_2$ contacts and transistor dimensions. The contact transfer length is ~35 nm at 300 K, which is verified experimentally using devices with 20 nm contacts and 70 nm contact pitch (CP), equivalent to the "14 nm" technology node.

**KEYWORDS:** MoS$_2$, contact resistance, 2D materials, transfer length, scaling, contact pitch

In order to achieve field-effect transistor (FET) gate lengths below 10 nm, transistor channel thicknesses below approximately 2 nm are required to maintain good gate control of the channel and to minimize leakage current.[1,2] In such thin films, 3-dimensional (3D) semiconductors like Si suffer from surface roughness (SR) effects: SR scattering reduces their mobility by nearly two orders of magnitude[3-8] and SR fluctuations (coupled with increases of band gap due to quantization effects[9]) can lead to strong variability in threshold voltage.[10,11] In contrast, when sufficiently clean, 2-dimensional (2D) materials such as graphene and $MoS_2$ do not have surface roughness and exhibit good electrical mobility that is largely independent of channel thickness.[12-14] $MoS_2$, a transition metal dichalcogenide (TMD), is a semiconductor with good mobility (~100 $cm^2V^{-1}s^{-1}$ in sub-2 nm thick films) and high on/off FET current ratio (~$10^7$) near room temperature.[14-20] Considering the nascent stage of ultra-thin TMD material synthesis and device fabrication, there is great potential for improvement, particularly from the perspective of device transport.

Nevertheless, despite the apparently robust *intrinsic* properties of 2D devices with respect to scaling, the contact resistance ($R_C$) currently limits further progress. As the channel length ($L$) is scaled down, the relative contribution of $R_C$ grows to dominate the total device resistance ($R_{TOT}$), eventually limiting performance. In addition, with overall transistor scaling, contact dimensions (like contact length $L_C$ in Figure 1a) must also be decreased, resulting in current crowding at the TMD-metal interface and further $R_C$ increases. To date, some improvements have been shown to $R_C$ of TMDs through work function engineering and doping,[21-26] however doping techniques frequently affect the threshold voltage, yielding devices which are difficult to turn off.

In this work we present a thorough study of contact resistance to $MoS_2$ under carefully controlled process conditions with various metals, combined with detailed modeling. We uncover that depositing Au contacts in ultra-high vacuum ($10^{-9}$ Torr) decreases $R_C$ down to ~740 $\Omega\cdot\mu m$ at room temperature, yielding long-term (>4 months) air stable-contacts without doping. This corresponds to a relatively small contact resistivity ($\rho_C \approx 3\times10^{-7}$ $\Omega\cdot cm^2$) and transfer length ($L_T \approx 30$ to 40 nm). These measurements also reveal that the transfer length method (TLM) approach gives a better estimate of contact resistance than four-probe measurements. Using our Au contacts we demonstrate the smallest $MoS_2$ FETs measured to date, with 20 nm contacts and 70 nm contact pitch (equivalent to the modern "14 nm" technology node), whose behavior confirms the transfer length estimated from TLM measurements.

**Experimental Results.** Thin MoS$_2$ flakes (1–15 layers) are exfoliated onto 90 nm of SiO$_2$ with a Si ($p^{++}$) substrate serving as the global back-gate (Figure 1a). Following XeF$_2$ etching to form well-defined channels, TLM structures[27] with varying channel lengths ($L$ = 0.1–3 µm) are defined by electron beam (e-beam) lithography and MoS$_2$ thicknesses are subsequently confirmed by atomic force microscopy (AFM), as shown in Figure 1b and further described in the Supporting Information Section 1. We e-beam evaporate various contacts (Ni, Ti, Au) under two deposition pressures ($P_D$ = 10$^{-9}$ and 10$^{-6}$ Torr) to examine the effects of metal type and fabrication on $R_C$. Estimated pressures during deposition are approximate, as they typically range from 0.5–5×10$^{-9}$ Torr for high vacuum and 0.5–5×10$^{-6}$ Torr for low vacuum. Devices are annealed at 300 °C for 2 hours and then measured, without further exposure to ambient, in the same vacuum probe station. Thermal annealing removes hysteresis and stabilizes electrical measurements, although it can sometimes slightly degrade Ni contacts (Supporting Information Section 2).

Figure 1c shows measured drain current ($I_D$) vs. gate voltage ($V_G$) data for Au-contacted MoS$_2$ ($P_D$ = 10$^{-9}$ Torr; thickness $d \approx$ 4.5 nm) with different channel lengths. The carrier density is obtained from the gate overdrive $V_G - V_T$, and the threshold voltage $V_T$ is deduced from the linear fit (red lines, Figure 1c) to the $I_D$ vs. $V_G$ curve at maximum transconductance,[27] $g_m = \partial I_D/\partial V_G$. All electrical data used in our analysis showed stable $V_T$ between measurements and no evident hysteresis. We specifically choose to determine contact resistance with the TLM approach rather than four-probe configurations which can introduce measurement errors,[27-29] and we provide a full comparison of the two methods for MoS$_2$ contacts in the Supporting Information Section 3.

Figure 1d shows good linear fits to the total device resistance normalized by width ($R_{TOT}$) vs. $L$, demonstrating uniform contacts. The vertical intercept of the linear fit yields the total contact resistance (2$R_C$) and the slope yields the intrinsic sheet resistance and mobility, for different carrier densities.[27] The inset of Figure 1d shows the extracted contact resistance vs. carrier density, including error bars arising from uncertainty of the linear extrapolation. The carrier density is $n = (V_G - V_T)C_{ox}/q$, where $C_{ox} \approx$ 38.4 nF/cm$^2$ for the SiO$_2$ employed here (90 nm) and $q$ is the elementary charge. This approach enables us to obtain the contact resistance $R_C$ for various carrier densities, different contact metals, temperatures and deposition conditions. As a result, Figure 1e shows $R_C$ vs. $n$ for Au, Ni, Ti contacts from this work ($d \approx$ 4–5 nm) and Au, Sc from other studies (d $\approx$ 6–8 nm).[21, 22] Our Au contacts with $P_D$ = 10$^{-9}$ Torr have the best quality, reaching 800 ± 200 Ω·μm (123 ± 30 Ω for $W$ = 6.5 μm) at $10^{13}$ cm$^{-2}$ carrier density, including the lead resistance. After subtracting the estimated lead contribution, the contact resistance is $R_C \approx 740$ Ω·μm at ~$10^{13}$ cm$^{-2}$ carrier density. These contacts have been air-stable for over 4 months without the oxygen sensitivity of Ni, and in particular that of low work function metals such as Sc and Ti (see Supporting Information Section 4).

Interestingly, as Figure 1e shows, Au contacts to samples of similar thickness but evaporated at higher pressure ($P_D$ = $10^{-6}$ Torr) exhibit three times higher $R_C$, both here and in ref. [21]. It is also important to note that our Ni and Au contacts with $P_D$ = $10^{-6}$ Torr are almost identical, and the Sc (ref. [22]) and our Au contacts with $P_D$ = $10^{-9}$ Torr (lowest pressure) have the lowest $R_C$. These improvements in $R_C$ with lower $P_D$ indicate that a cleaner metal-MoS$_2$ interface is crucial for better contacts, almost regardless of bulk metal work function. We also emphasize that our use of multiple (> 6) TLM channel lengths down to 100 nm is important for accurately determining the contact resistance, because fewer and longer channel lengths can lead to significantly larger errors in the extracted $R_C$ (see Supporting Information Section 5).

**Modeling Analysis.** To understand the improvement in $R_C$ with lower $P_D$, we first examine the effective contact barrier height, $\phi_{eff}$, in Figure 2a. The current density of thermionic emission through a metal-semiconductor contact[27, 30] is

$$J = A^* T^2 e^{-\frac{q\phi_{eff}}{kT}} (e^{\frac{qV}{kT}} - 1) \qquad (1)$$

where $A^*$ is the Richardson constant, $V$ is the applied bias, $T$ is the temperature, $k$ is Boltzmann's constant. Using this equation, the slope of the Richardson plot, $\ln(I_D/T^2)$ vs. $1/T$, yields $\phi_{eff}$ as a function of the vertical electric field ($E_N$) between the gate and channel. $\phi_{eff}$ represents both thermionic and field emission through the Schottky barrier and can change with $E_N$, which alters the barrier width. In general, the Schottky barrier height $\Phi_B \neq \phi_{eff}$, but an increase in $\phi_{eff}$ corresponds to an increase in $\Phi_B$ for constant carrier density (see Supporting Information Section 6). Figure 2a displays $\phi_{eff}$ for Au contacts with $P_D$ = $10^{-6}$ and $10^{-9}$ Torr, respectively. Both types of contacts show similar $\phi_{eff}$ values over the range of $E_N$, indicating that a change in $P_D$ has little effect on the barrier height. Thus an improved $P_D$ appears more likely to affect $R_C$ through the lateral access resistance, not the interface resistance, which is determined by $\Phi_B$.

The $R_C$ measured by TLM is a total contact resistance resulting from two primary contributions: 1) thermionic and field emission through the metal-MoS$_2$ Schottky barrier and 2) lateral access resistance under the contact due to the sheet resistance ($R_{SH}$) of MoS$_2$. To understand the importance of each contribution, we calculate $R_C$ vs. $T$ using a model that accounts for both the interfacial resistivity and the access resistance. First, we use a Tsu-Esaki[31, 32] model with a transfer matrix method to calculate the specific interfacial resistivity ($\rho_i$) of the metal-MoS$_2$ interface including both tunneling and thermionic emission. $\rho_i$ only accounts for transport through the Schottky barrier whereas the specific contact resistivity ($\rho_C$), discussed shortly, includes both the Schottky barrier and interlayer transport underneath the contacts. Additional model details are provided in the Supporting Information Section 7. Second, the sheet resistance is calculated from the mobility of MoS$_2$ under the contact, $\mu_C$. We take the temperature dependence $R_{SH}(T) = [qn\mu_C(T)]^{-1}$ to be dictated by the measured $T$ dependence of mobility, $\mu_C(T) = \mu_{C0}(T/300)^{-1.6}$ (Figure 2b) where $\mu_{C0}$ is the mobility at 300 K. In doing this, we allow the mobility under the contact to differ from that in the channel ($\mu$), with $\mu_{C0}$ as a fitting parameter ($\mu_{C0}$ = 35 cm$^2$V$^{-1}$s$^{-1}$ in Figure 2c).

The total contact resistance $R_C$ can either increase or decrease with $T$ depending on the relative contributions of $R_{SH}$ (increasing with $T$ due to phonon scattering) and of $\rho_i$ (decreasing with $T$ due to thermionic emission), as summarized in Figure 2c. Thus, examining the temperature dependence of $R_C$ can illuminate the physical mechanisms (interface vs. access resistance) limiting the current transport at MoS$_2$-metal contacts. We use the well-known transmission line model (Figure 2b inset)[27, 33] to account for both $\rho_i$ and $R_{SH}$ in the $R_C$ calculation:

$$R_C = \frac{\rho_C}{L_T}\coth\left(\frac{L_C}{L_T}\right) \approx \sqrt{\rho_i R_{SH}} \qquad (2)$$

where $R_C$ is in units of Ω·μm, normalized by the contact width $W$ for easier comparison between devices of different widths. $L_T$ is the current transfer length (Figure 3b inset), $L_C$ is the physical contact length (Figure 1a or Figure 3b inset), and the approximation above holds if $L_T \ll L_C$ (= 500 nm here) and $\rho_C \approx \rho_i$, which we will show are both reasonable assumptions.

With this in mind, Figure 2d shows our model calculations of $R_C(T)$ for different barrier heights, assuming the mobility in the channel and under the contacts are the same, $\mu_C = \mu$. As $T$ increases, thermionic emission initially causes a decrease in $R_C$. At $T >$ 200 K, $R_C$ eventually be-comes either constant or rises slightly with $T$ due to the increase of phonon-limited access resistance. The latter effect is more pronounced if $\mu_C < \mu$, as shown in Figure 2e and discussed below. Figure 2f reveals the measured temperature dependence of the regular Ni and the cleanest Au contacts, which can be best fitted using our model with $\Phi_B = 150$ meV (consistent with previous measurements for Ni and Au contacts[22, 34]) and $\mu_{C0} = 0.25$ and 20 cm$^2$V$^{-1}$s$^{-1}$, respectively. This suggests that the improved $\mu_C$ with clean Au contacts is correlated with the lower $P_D$, which reduces the amount of adsorbates (impurities) trapped at the metal-MoS$_2$ interface during the contact deposition. This hypothesis is also supported by colorized scanning electron microscope (SEM) images of our contacts, which reveal the grain structure of the various metals on MoS$_2$ (see Supporting Information Section 8).

Our analysis suggests that $R_C$ is controlled more by the lateral access resistance under the contact rather than the Schottky barrier. However, given that $R_{SH}$ depends both on $\mu$ and $n$, we cannot rule out the possibility that $R_C$ is affected by $n$ under the contact instead of $\mu$ (e.g. through charge depletion, which in turn is microscopically influenced by the deposition conditions). This scenario would also lead to a larger $R_{SH}$ of MoS$_2$ under the contacts, although it would have a weaker temperature dependence. Our model and measurements cannot presently distinguish between the two scenarios, but future work with higher contact doping and a two-dimensional solution of the field, charge and current distributions at the contacts could help elucidate this issue.

**Current Transfer Length.** While calculating $\rho_i$ is useful for modeling purposes, $\rho_C$ can be directly extracted using the transmission line model in conjunction with the TLM measurement because $R_C$ and $R_{SH}$ are known a priori. $R_{TOT}$ vs. $L$ (Figure 1d) yields the sheet resistance (slope of the line fits) and the contact resistance (vertical intercept = $2R_C$), yielding both $\rho_C$ and $L_T$ from the exact form of eq. (2) and from $L_T = (\rho_C/R_{SH})^{1/2}$. Thus, we extract $\rho_C$ (Figure 3a) and the transfer length $L_T$ (Figure 3b) from $R_C$ measurements on Au contacts ($P_D = 10^{-9}$ Torr) for varying $T$ and $n$. At 300 K the lowest effective contact resistivity is $\rho_C \approx 3 \times 10^{-7}$ Ω·cm$^2$ for our cleanest Au-MoS$_2$ contacts; this result is comparable to that of chemically doped contacts, but without the disadvantage of a highly doped, "always on" channel with very negative $V_T$. $\rho_C$ decreases with increasing $T$ due to increased thermionic emission, down to $1.5 \times 10^{-7}$ Ω·cm$^2$ at 400 K. Despite these improvements, $\rho_C$ remains higher than for modern Si contacts (~$10^{-8}$ Ω·cm$^2$),[35] indicating a continued need for improvement. Note that $\rho_C$ is slightly larger than $\rho_i$ (see Supporting Figure

S10) due to vertical, intra-layer resistance underneath the contact, which is not taken into account with the Tsu-Esaki model.

In Figure 3b we extract $L_T \approx 40$ nm at 300 K, a relatively small value consistent with a large $R_{SH}$ and small $\rho_C$. The transfer length rapidly decreases with rising temperature, as $R_{SH}$ increases due to phonon scattering and $\rho_C$ decreases from enhanced thermionic emission. In this respect, MoS$_2$ transistors look promising from the point of view of contacts at elevated temperatures ($L_T \approx 20$ nm at 400 K), but more improvements must made at room temperature and below.

**Transistor Scaling.** We now turn to an assessment of the scaling limits of MoS$_2$ FETs. In Figure 4a, using our best Au contacts, we represent both the intrinsic channel resistance ($R_{CH}$) and the total contact resistance ($2R_C$) as a fraction of the total device resistance ($R_{TOT} = 2R_C + R_{CH}$) for channel lengths $L \leq 1$ µm. At 300 K, the contact resistance does not dominate ($R_{CH} \geq 2R_C$) down to $L \approx 90$ nm, where the two resistance components are approximately equal. In other words, the channel length cannot be scaled below 90 nm given the best techniques shown in the present work, at room temperature. At 80 K, reduced phonon scattering results in a higher mobility (here $\mu \approx 140$ cm$^2$V$^{-1}$s$^{-1}$) and a less resistive channel, thus MoS$_2$ FETs become contact limited below 0.6 µm. However, above room temperature (here at 400 K) the scalability of MoS$_2$ devices is improved, as the contact resistance does not become dominant until below $L \approx 40$ nm.

These comparisons place the need for "good" contact resistance in the proper context. In other words, $R_C$ ought to be evaluated as a fraction of the total device resistance including the channel. If the channel mobility is improved, the contact resistance must also be reduced for a given channel length. (For this reason, graphene FETs have more stringent contact resistance requirements than MoS$_2$ FETs.) Conversely, devices with lower mobility may be scalable to smaller channel lengths for a given contact resistance. These issues are particularly important for MoS$_2$, where $R_{CH}$ and $R_C$ have opposite temperature dependencies, as seen in Figures 2c and 4a.

For completeness, it is also important to account for the contact length $L_C$, which ultimately plays a large role in determining the total device size and density. The contact pitch[36, 37] (CP) is $L + L_C$, taking $L$ as the inner source-to-drain spacing (Supporting Figure S11). The CP is the true measure of device density for a given transistor technology.[36-39] We recall that $R_C$ is independent of $L_C$ for $L_C \gg L_T$, but $R_C$ increases sharply as a result of current crowding for $L_C < L_T$. (A more complete scaling analysis must also take into account the scaling of gate-to-contact spacers and that of parasitic capacitances with $L_C$.[36,37]) Thus, the best Au contacts reported here suggest a lower limit of $L_C$ ~ 40 nm at room temperature and ~20 nm at 400 K (see Figure 3b), before current crowding begins to play a role at the contacts.

To investigate these points, we also fabricated MoS$_2$ FETs with $L_C$ down to 20 nm for the first time. Figure 5a shows the output characteristics of MoS$_2$ FETs with varying contact lengths fabricated on the same MoS$_2$ flake (of 2-3 layers). As expected, no current degradation results from decreasing $L_C$ = 250 nm to 100 nm, since $L_C \gg L_T$. However, decreasing $L_C$ from 100 nm to 20 nm degrades the current by 30%. Applying a fit to $I_D$ vs. $L_C$ (Figure 5a, inset) using equation 2 yields an estimated $L_T \approx$ 30 nm, which is consistent with the extracted value of $L_T \approx$ 40 nm from TLM measurements. Although the 20 nm contacts are smaller than the transfer length, $I_D$ vs. $V_D$ measurements nevertheless yield drive currents greater than 300 µA/µm (Figure 5b). TEM cross-sections (Figure 5c) confirm a contact pitch of 70 nm for the smallest fabricated device, approximately corresponding to the modern "14 nm" technology node.[40] To the best of our knowledge, these represent the smallest TMD FETs fabricated to date. In addition, the high drive currents demonstrated (>300 µA/µm) are also a record for a TMD FET at these dimensions. Nonetheless, proper current saturation is not observed because these highly scaled devices are contact-limited, emphasizing the need for further improvements to nanoscale TMD contacts.

To enable well-behaved MoS$_2$ FETs with sub-10 nm gate lengths at room temperature, $R_C$ remains to be decreased by at least an order of magnitude. As we have shown, the lateral transport under the contact (access resistance) can be more important than $\Phi_B$ for improving $R_C$ to MoS$_2$. Further improvements to the access resistance in MoS$_2$ FETs will result from improving the carrier density under the contacts, particularly through doping.[23-25] To understand this effect, in Figure 4b we compare measured and calculated $R_C(T)$ for various carrier densities. $R_C$ decreases significantly as $n$ rises up to ~5×10$^{12}$ cm$^{-2}$ due to increased thermionic and field emission through the Schottky barrier, while at higher carrier densities $R_C$ is limited by $R_{SH}$. To achieve a desirable $R_C \approx$ 100 Ω·µm (consistent with current ITRS requirements[35]), significantly higher doping ($n > 3 \times 10^{13}$ cm$^{-2}$) is required to decrease $R_{SH}$ under the contact. This value gives an important target to be pursued by chemical or molecular doping.

To put 2D devices and materials in perspective, Figures 6a and 6b compare our results (and others from the literature) for contact resistance and mobility with those of Si technology of similar channel thickness, $d$. Interestingly, our "clean" Au contacts ($R_C \sim 740$ Ω·μm for $d \approx 4.5$ nm) are comparable to those of Si FinFETs of similar thickness,[41, 42] which also display large access resistance. Ultra-thin Si fins suffer both from mobility degradation (described in our introduction) and from difficulty in doping and siliciding such thin Si layers. Our monolayer ($d \approx 0.65$ nm) $MoS_2$ devices typically show higher $R_C$ due to greater access resistance, ostensibly from carrier-substrate scattering. However, if this mobility degradation can be mitigated, properly doped monolayer $MoS_2$ contacts could be superior to those of Si in sub-nanometer thickness films.

In Figure 6b we compare mobilities of 2D materials to those of 3D materials like Si in ultra-thin body Si-on-insulator (UTB-SOI) as a function channel thickness. As previously mentioned, UTB-SOI and FinFETs are hampered by surface roughness scattering, causing the mobility to decrease rapidly for $d < 3$ nm.[3-8] (A similar behavior is also seen in other ultra-thin 3D semiconductors like SiGe and III-Vs.[43, 44]) By comparison, $MoS_2$ and graphene mobility remain relatively constant down to sub-1 nm monolayer thickness, a unique characteristic of 2D materials, and further improvements are expected given the nascent stages of these technologies. Hole mobilities of UTB-SOI are even lower in thin channels (<20 $cm^2V^{-1}s^{-1}$ for $d = 2.5$ nm),[3] whereas those of sub-1 nm 2D semiconductors such as $WSe_2$ appear to be reasonable (>200 $cm^2V^{-1}s^{-1}$).[45] The preservation or enhancement of mobility, in tandem with further lowering of contact resistance in 1-3 layer 2D devices should thus remain a key focus in the 2D research community.

**Summary:** We presented a comprehensive study of contact resistance to $MoS_2$ FETs under carefully controlled process conditions with various metals, temperatures, and detailed modeling. TLM structures were preferred (instead of four-probe) as we have found they are more reliable for $R_C$ measurements. We also found that the use of multiple (> 6) TLM channel lengths down to 100 nm is important for accurately determining the contact resistance. Combining modeling and experiments, we separated the two components of the contact resistance, i.e. "lateral" access resistance under the contact in series with "vertical" transport across the Schottky barrier (and across $MoS_2$ layers in multilayer devices). The lateral access resistance dominates, and appears to be the key component which must be decreased in future efforts.

We uncovered that ultra-high vacuum Au deposition provides a higher quality metal-MoS$_2$ interface, leading to $R_C$ as low as 740 Ω·μm at room temperature without deliberate doping, and therefore stable for over 4 months. The estimated contact resistivity ($\rho_C \approx 3\times10^{-7}$ Ω·cm$^2$) and transfer length ($L_T \approx 35$ nm) must be assessed in the proper context of transistor pitch and density. As a demonstration, we fabricated MoS$_2$ transistors with 20 nm contacts and 70 nm contact pitch, corresponding to the "14 nm" technology node. We also discussed further advancements that must occur if scaling of 2D-FETs below 10 nm is desired.

## ■ ASSOCIATED CONTENT

**Supporting Information**

Details of MoS$_2$ device fabrication and characterization; thermal annealing of devices and contacts; four-probe vs. transfer length method measurements; contact (resistance to) degradation, metal-MoS$_2$ interface resistance model; Schottky barrier height extractions; contact morphology. Material is available free of charge via the Internet at http://pubs.acs.org.

## ■ ACKNOWLEDGMENT

This work was supported in part by the Air Force Office of Scientific Research (AFOSR) FA9550-14-1-0251, the National Science Foundation (NSF) EFRI 2-DARE program 1542883, and the Stanford SystemX Alliance. C.D.E. acknowledges support from a Stanford Graduate Fellowship (SGF).

## ■ REFERENCES

■ **FIGURES**

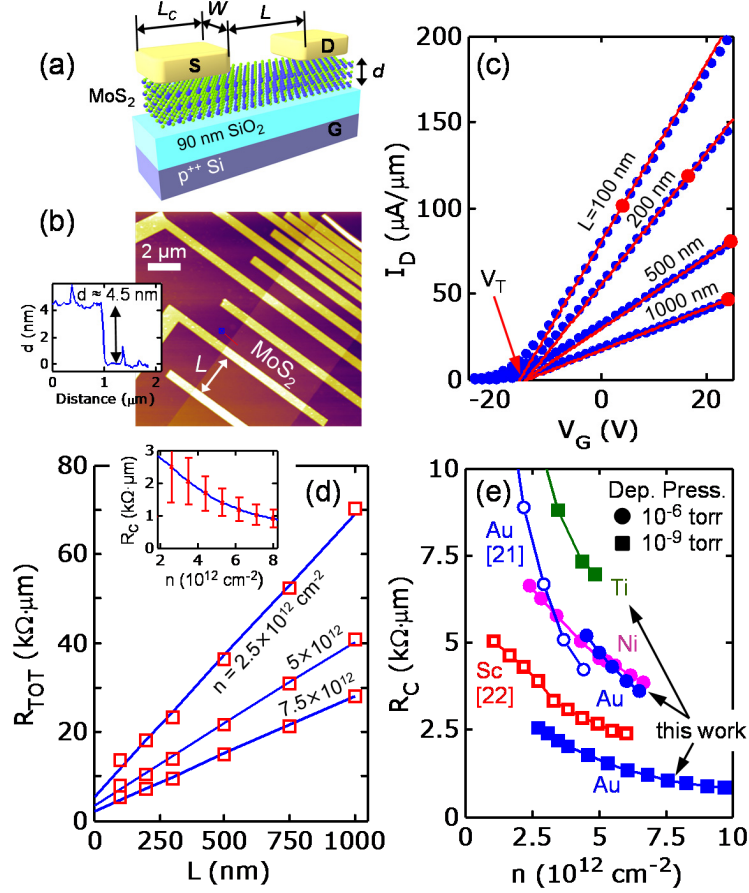

**Figure 1:** (a) Schematic of our MoS$_2$ devices. (b) AFM image of TLM structure on MoS$_2$. Inset: AFM cross-section of height profile. (c) Measured current vs. gate voltage ($V_D$ = 1 V) for Au electrodes deposited at 10$^{-9}$ Torr, showing $V_T$ extraction. (d) Total device resistance $R_{TOT}$ vs. $L$ measured by TLM, at various carrier densities, $n$. Linear extrapolation of $R_{TOT}$ vs. $L$ yields $2R_C$ as the vertical axis intercept. Inset: $R_C$ vs. $n$ for our "clean" Au contacts. (e) Measured $R_C$ vs. $n$ for multiple contact metals at different deposition pressures, from this and previous work.[21, 22] Lower deposition pressures lead to cleaner interfaces and lower $R_C$. The cleanest Au contacts used here reach $R_C \approx$ 740 Ω·μm at $n \approx 10^{13}$ cm$^{-2}$ after the metal lead resistance is subtracted.

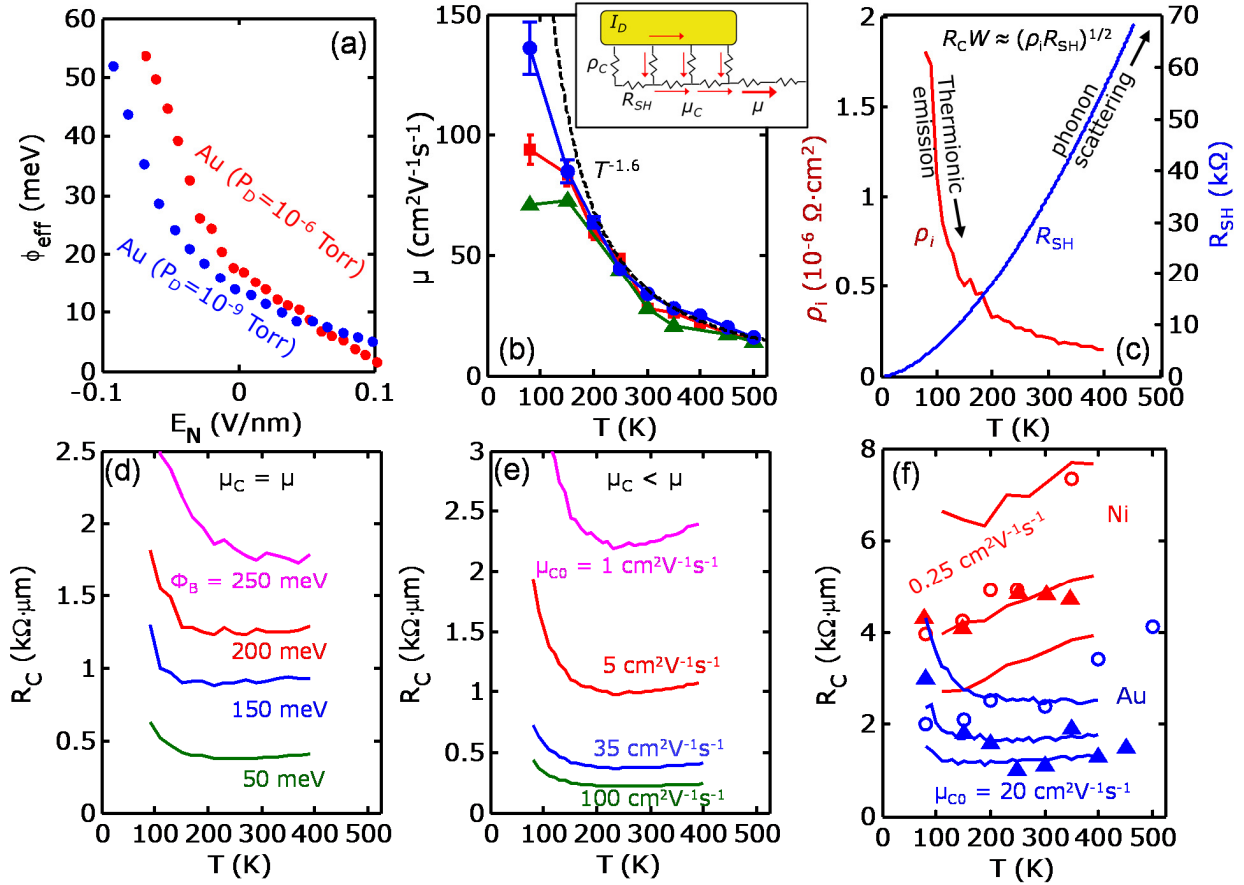

**Figure 2:** (a) $\phi_{eff}$ vs. $E_N$ (normal gate electric field) for Au contacts with $10^{-6}$ Torr (red) and $10^{-9}$ Torr (blue) deposition pressure. The uncertainty of both data sets is ±10 meV. (b) Measured intrinsic mobility $\mu$ vs. $T$ for MoS$_2$ with Au (blue) and Ni (red/green) contacts, showing $T^{-1.6}$ dependence. Inset: Schematic of contact as a resistor network, highlighting the different $\mu_C$ and $\mu$. (c) Calculated temperature dependence of sheet resistance $R_{SH}$ (blue) and interfacial resistance $\rho_i$ (red) at $n = 5\times10^{12}$ cm$^{-2}$. (d) Calculated $R_C(T)$ for $\mu_C = \mu$, $n = 5\times10^{12}$ cm$^{-2}$, and varying $\Phi_B$ as listed. (e) Calculated $R_C(T)$ for varying $\mu_C$, $\Phi_B = 150$ meV and $n = 5\times10^{12}$ cm$^{-2}$. (f) Measurements of $R_C$ vs. $T$ for Au ($10^{-9}$ Torr, blue symbols) and Ni ($10^{-6}$ Torr, red symbols) contacts at $n = 5\times10^{12}$ cm$^{-2}$. Calculations (solid lines) are shown for $\mu_{C0} = 0.25$ cm$^2$V$^{-1}$s$^{-1}$ (red) and 20 cm$^2$V$^{-1}$s$^{-1}$ (blue) for $n = 2.2$, 4.3, and $5.5\times10^{12}$ cm$^{-2}$ from top to bottom. Empty and filled symbols represent two different samples.

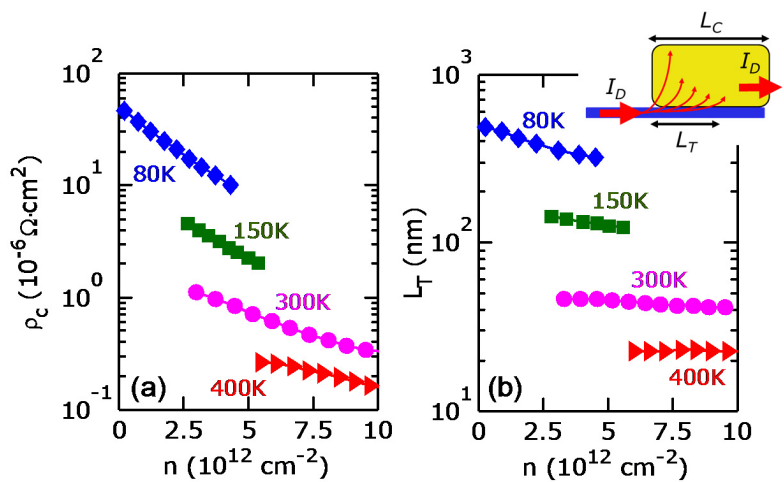

**Figure 3:** Specific contact resistivity $\rho_C$ and transfer length $L_T$ as a function of carrier density and temperature for our "clean" Au contacts ($10^{-9}$ Torr deposition). (a) $\rho_C$ vs. $n$ obtained directly from the TLM measurement. The corresponding interfacial and contact resistivity are shown in Supplementary Figure S10. (b) $L_T$ vs. $n$ for $T = 80$–$400$ K. Inset schematic shows $L_T$ and the physical contact length $L_C$; red arrows are current flow lines. The extracted $L_T$ uncertainty is approximately 33% at high carrier density and 50% at low carrier density.

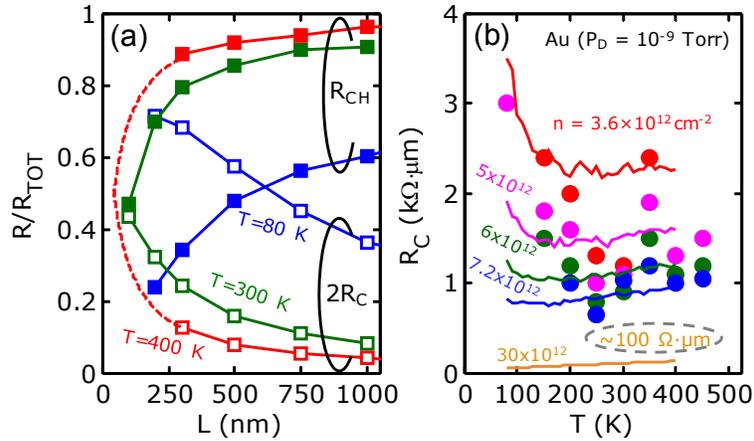

**Figure 4:** (a) Fraction of channel resistance ($R_{CH}$) and contact resistance ($2R_C$) contributing to the total device resistance ($R_{TOT} = R_{CH} + 2R_C$) as a function of channel length $L$, at three temperatures, for our "best" measured contacts. $2R_C$ dominates $R_{TOT}$ for $L \leq 40$ nm at 400 K, $L \leq 90$ nm at 300 K and $L \leq 600$ nm at 80 K. Solid symbols are our measured TLM data. Dashed line is a simple extrapolation based on the known $R_{SH}$ and $R_C$. (b) $R_C$ vs. $T$ for different carrier densities. Symbols connected are experimental data; solid lines are calculations with $\Phi_B = 150$ meV and $\mu_C = 20$ cm$^2$V$^{-1}$s$^{-1}$ corresponding to the same $n$ as the measurements, where data are available. For the highest doping ($3\times10^{13}$ cm$^2$) calculations suggest $R_C \approx 100$ Ω·μm at 300 K.

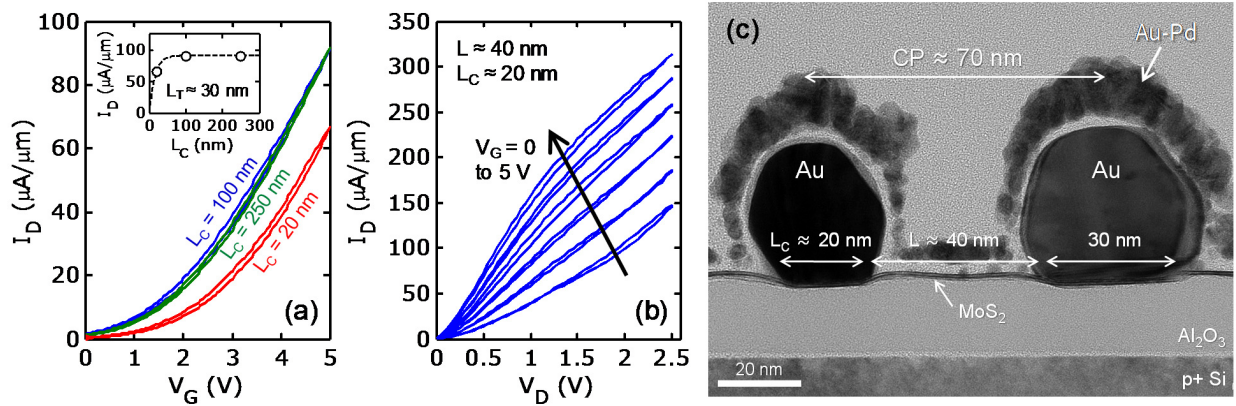

**Figure 5:** (a) Measured current vs. gate voltage for "short" Au contacts with $L_C$ = 20, 100, 250 nm and channel $L$ = 40 nm (at $V_D$ = 1 V). Inset: Measured current (circles), simulated current (dashed line) vs. contact length at $V_G$ = 5 V, yielding a transfer length $L_T \approx 30$ nm. (b) Measured current vs. $V_D$ for the smallest device measured ($L_C \approx 20$ nm) showing $I_D > 300$ µA/µm, a record for a TMD FET at ~70 nm contact pitch. Two sweeps for each data set reveal minimal hysteresis. (c) TEM cross-section of a MoS$_2$ FET with nanoscale contacts. The residue covering the device is sputtered Au-Pd, applied during preparation for the TEM cross-section. The gate is the p$^+$ Si seen below the Al$_2$O$_3$ dielectric.

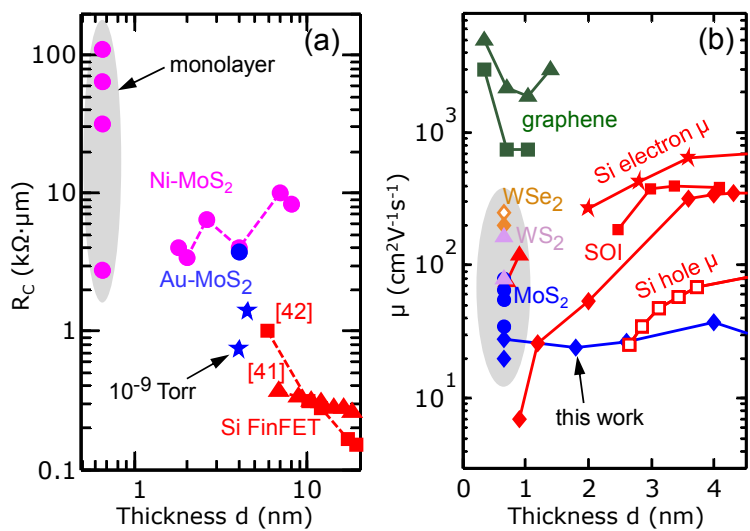

**Figure 6:** Scaling of contact resistance and mobility with film thickness *d* for 2D materials (on SiO$_2$) and ultra-thin Si fins or films, at room temperature. (a) $R_C$ vs. *d* for our Ni-MoS$_2$ (magenta), our Au-MoS$_2$ contacts (blue), and Si FinFET contacts (red) from the literature.[41, 42] Our Ni and Au contacts marked by circles were deposited at 10$^{-6}$ Torr, the stars at 10$^{-9}$ Torr. All our data are based on TLM, except the monolayers (see Supporting Information Section 11), and correspond to the highest carrier density for each sample ($5 \times 10^{12}$ to $10^{13}$ cm$^{-2}$). (b) $\mu$ vs. *d* for graphene (green),[12, 13] MoS$_2$ (blue, including this work)[46-49], WSe$_2$ (orange)[45, 50], WS$_2$ (purple)[51], and ultra-thin SOI at $5 \times 10^{12}$ cm$^{-2}$ carrier density (red).[3-6] Closed (open) symbols represent electron (hole) mobilities. Mobilities here are for films in contact with SiO$_2$. Some values are for field-effect and others for effective mobility,[27] thus comparisons are approximate.

# Supporting Information
## Improved Contacts to MoS$_2$ Field-Effect Transistors by Ultra-High Vacuum Metal Deposition

Chris D. English[1], Gautam Shine[1], Vincent E. Dorgan[2], Krishna C. Saraswat[1], Eric Pop[1]

[1] Electrical Engineering, Stanford University, Stanford, CA, 94305, USA
[2] Electrical & Computer Eng., Univ. Illinois Urbana-Champaign, Urbana, IL, 61801, USA

## 1. MoS$_2$ Device Fabrication and Characterization

We exfoliate MoS$_2$ flakes onto 90 nm of SiO$_2$ supported by a highly doped Si substrate (p-type, resistivity < 5x10$^{-3}$ Ω·cm) using the "tape method." Monolayer and multilayer flakes are first identified with optical microscopy (Figure S1a, showing flake in a completed TLM device), and then confirmed with Raman spectroscopy (Figure S1b). We then use atomic force microscopy (AFM) to verify the flake thicknesses (Figure S1c-d). Before depositing the metal electrodes we use electron beam (e-beam) lithography and a XeF$_2$ etch to pattern the MoS$_2$ into a well-defined channel. The XeF$_2$ etch consists of 2 cycles at 60 s/cycle and XeF$_2$ pressure = 3 Torr. Finally, we use e-beam lithography again to pattern the metal electrodes. Two different e-beam evaporation systems are used in the metal depositions. The first one has a base pressure of ~10$^{-6}$ Torr, while the other has a base pressure of <10$^{-9}$ Torr. The high-vacuum system remains at low pressure for 3-month-long intervals and the sources are rarely changed, maintaining cleanliness. Using an ultra-high vacuum system ($P_D = 10^{-9}$ Torr) for metal deposition is particularly important for low-workfunction metals such as Sc which may oxidize very easily.

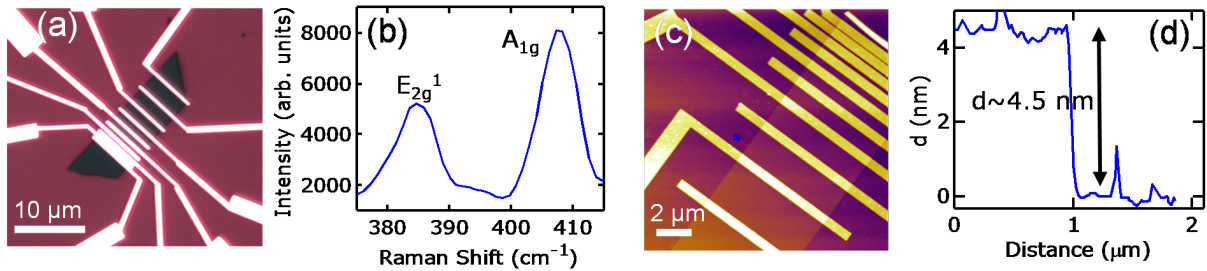

**Fig. S1:** (a) Optical image of a typical TLM structure on exfoliated MoS$_2$. (b) Raman spectrum of a MoS$_2$ device showing the characteristic E$_{2g}^1$ and A$_{1g}$ peaks. (c) AFM image of a TLM structure on exfoliated MoS$_2$. (d) Height profile of the MoS$_2$ flake by AFM, along the red cut-line in (c).

## 2. Thermal Annealing of Devices and Contacts

Typically, devices are annealed in our vacuum probe station (~10$^{-5}$ Torr) for 2 hours at $T$ = 300 °C just before electrical measurements are performed without breaking vacuum. The effect is to evaporate adsorbates off the surface of the MoS$_2$ channel, resulting in substantially reduced hysteresis and improved current drive. Note that the improved current drive results from a decrease in $V_T$ in the channel, not from an improvement in $R_C$. Vacuum annealing can have a small detrimental effect on $R_C$ as shown in Figure S2.[*] For Ni contacts in particular, $R_C$ appears to increase by ~25% after the thermal anneal, however a lesser effect is noticed for the oxidation resistant Au contacts. Although we do not have similar measure-ments for Ti contacts, we believe they suffer similar degradation like Ni, due to their tendency to oxidize. Since the thermal annealing results in device stability and lack of hysteresis (which are important for consistent electrical measurements), it has an overall positive effect for our characterization.

---

[*] The term "vacuum annealing" could be more properly called "low-pressure annealing" as the probe station at 10$^{-5}$ Torr will still retain some residual O$_2$ and H$_2$O.

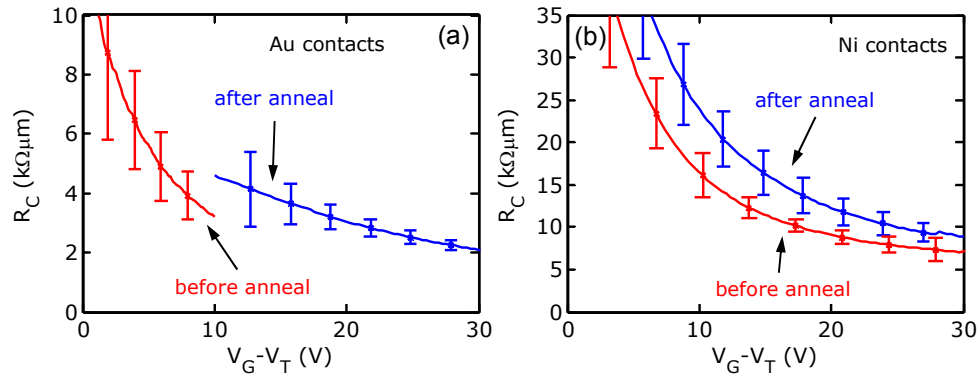

**Fig. S2:** $R_C$ for Au (a) and Ni (b) contacts extracted from TLM structures. Measurements performed immediately before (red) and immediately after (blue) the vacuum anneal, in the same probe station, without breaking vacuum. The different over-drive voltage ranges in Fig. S2a resulted from a decrease in $V_T$.

## 3. Four-probe vs. Transfer Length Method (TLM) Measurements

We find that four-probe measurements on TMDs can yield inaccurate estimates of $R_C$. To demonstrate this, we have measured the $R_C$ of Ni-MoS$_2$ contacts with a four-probe structure and a TLM structure on the same MoS$_2$ flake. In a four-probe configuration (Figure S3a), a bias current ($I_{BIAS}$) is applied between the outer electrodes, while the voltage on the inner sense electrodes is measured ($V_1$ and $V_2$). Ideally, the sense electrodes should not alter the current path in the material, allowing for an accurate measurement of the channel resistance by $R_{CH} = (V_1 - V_2)/I_{BIAS}$. However, the inner sense electrodes can be invasive, shunting the current through the path of least resistance as shown in Figure S3a. The shunted current causes an additional voltage drop across the Ni-MoS$_2$ interface caused by $R_C$, leading to an inaccurate measurement of the channel resistance in the four-probe configuration ($R_{CH}^*$) (Figure S3b). We account for this with a parameter $0 < \alpha < 1$, which indicates the portion of current shunted by the sense electrodes.

$$R_{CH}^* \approx R_{SH} + \alpha(2R_C)$$

$$R_C^* = (R_{TOT} - R_{CH}^*)/2 = (2R_C + R_{SH} - R_{SH} - \alpha(2R_C))/2 = R_C(1-\alpha)$$

$$\alpha = 1 - R_C^*/R_C$$

Here $R_C^*$ is the (inaccurate) contact resistance extracted by the four-probe measurement, while $R_C$ is the true contact resistance measured by TLM. In other words, *the four-probe measurement will always underestimate the contact resistance, due to the current shunting through the sense electrodes.* $R_{TOT}$ is the total device resistance found by a two-probe measurement on the inner sense electrodes in the four probe configuration. This simple model is not exact since $R_C$ includes both the interface and access resistance, but it illustrates the essential physics. In Figure S3c, $\alpha$ initially increases with $V_G$ due to the sharp decline in $\rho_C$. However, $\alpha$ eventually decreases as the sheet resistance under the contact decreases with $V_G$. Note that $\alpha \to 0$ as $R_{SH} \to 0$, and $\alpha \to 1$ as $R_{SH} \to \infty$. *With the four probe measurement used here, $R_C^*$ underestimates $R_C$ by more than a factor of 10* at $V_G-V_T = 10$ V (maximum $\alpha$), where $R_C^* \approx 1.5$ k$\Omega\cdot\mu$m and $R_C = 30$ k$\Omega\cdot\mu$m.

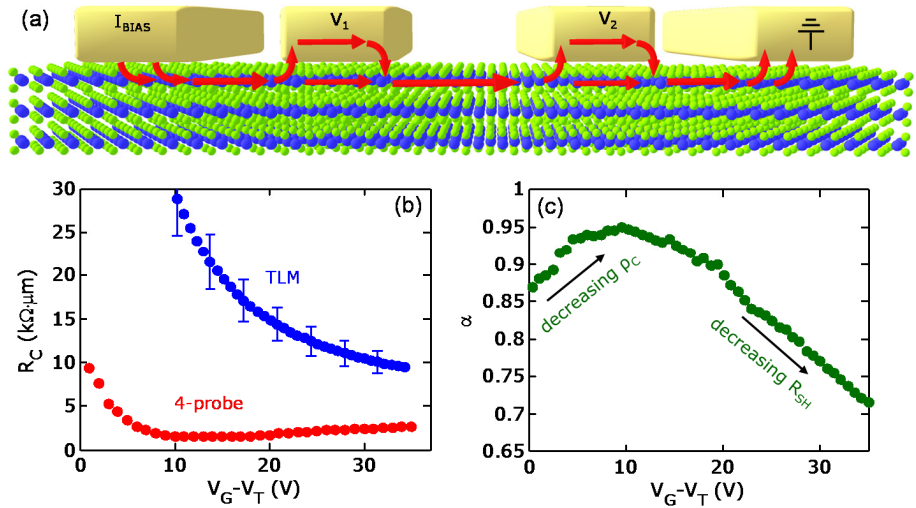

**Fig. S3:** (a) Representation of a typical four-probe measurement structure. A bias current ($I_{BIAS}$) is applied between outer electrodes, while the voltage is measured on the inner sense electrodes ($V_1$ and $V_2$). Red arrows indicate the direction of current flow. (b) Measured contact resistance ($R_C$) vs. overdrive voltage ($V_G$-$V_T$) for transfer length method (TLM) and four-probe measurements showing how much the four-probe measurements can *underestimate* $R_C$. (b) α vs. ($V_G$-$V_T$) for the graph in part (b). α = 1 − $R_C^*$/$R_C$. $R_C^*$ is the contact resistance obtained by the 4-probe measurement and $R_C$ is that measured by TLM.

## 4. Contact Degradation

Au contacts are resistant to interface degradation over time (Figure S4a) *without an encapsulation layer*, while Ni contacts degrade significantly over 6 months (Figure S4b). The degradation is likely due to oxidation at the interface. Low work function metals such as Sc and Ti also show significant $R_C$ degradation due to oxidation, even just after the contact deposition. Note that resistance to degradation over time, in this case, is solely related to the oxidation resistance of metal. Thus, Au contacts deposited at low or high vacuum should be resistant to degradation.

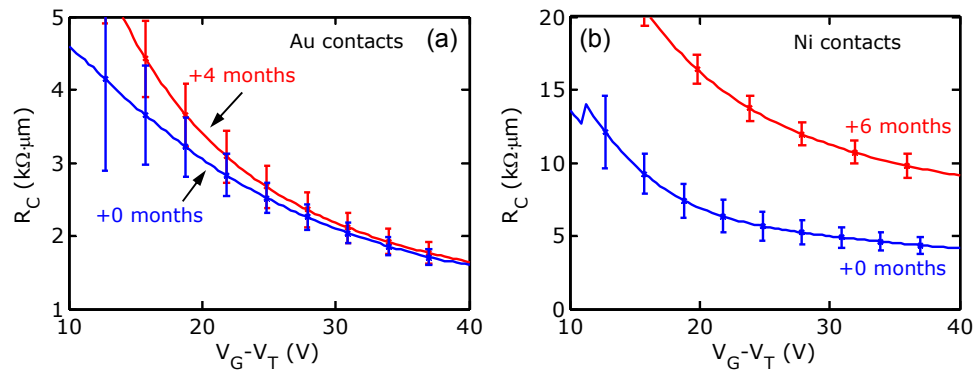

**Fig. S4:** $R_C$ for Au (a) and Ni (b) contacts, comparing the degradation of the $R_C$ with time.

We plot the *total* device resistance ($R_{TOT}$) to show the device degradation over time. Figure S5 shows both long channel ($R_{CH}$ dominated) and short channel ($R_C$ dominated) devices to decouple the effects of degradation on each part of the device (channel- vs. contact-dominated). For long channel devices with Ni and Au contacts, there is very little increase in $R_{TOT}$, indicating only a minor degradation of the channel. However, for the short channel devices, Au contacts remain the same while the Ni contacts show a large increase. The difference here indicates that the mechanism of contact degradation is related solely to the contact metals themselves, not to degradation of the $MoS_2$ mobility (from adsorbates, S vacancies, etc.).

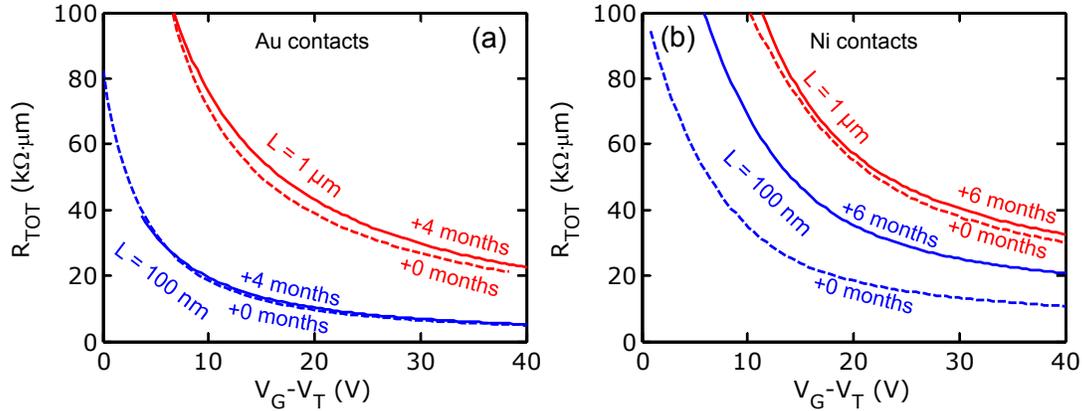

**Fig. S5:** (a) Total device resistance ($R_{TOT}$) vs. gate overdrive voltage ($V_G - V_T$) for both short ($L$ = 100 nm) and long ($L$ = 1 μm) channel devices using Au contacts deposited at ultra-high vacuum ($P_D = 10^{-9}$ Torr). (b) $R_{TOT}$ vs. ($V_G - V_T$) for both short ($L$ = 100 nm) and long ($L$ = 1 μm) channel devices using Ni contacts deposited at low vacuum ($P_D = 10^{-6}$ Torr). Measurements taken just after are indicated by the dashed lines. Measurements taken 4-6 months later are solid lines.

## 5. Accuracy of TLM Extractions

We point out the importance of using careful, accurate TLM measurements, and some potential pitfalls. Our TLM measurements are very robust due to the following implementations:

**1) We account for threshold voltage ($V_T$) variation between the different MoS$_2$ channels.** For instance, Figure S6a below shows the $V_T$ variation for a device with clean Au ($P_D = 10^{-9}$ Torr) contacts. The $V_T$ variation (-18 V to -22 V) is significant, and must be accounted for. We eliminate the $V_T$ variation by extracting $R_C$ at specific overdrive voltages ($V_G - V_T$), resulting in a more accurate estimation of $R_C$.

**2) We use a larger number (≥ 6) of channel lengths for improved $R_{TOT}$ vs $L$ fits, down to $L$ = 100 nm.** Figure S6b shows one of our TLM extractions ($R_{TOT}$ vs $L$) with deliberately fewer, only 4 channel lengths. Note that even though the 4 data points ($R_{TOT}$) appear to be very co-linear, the resulting $R_C$ estimation yields large errors ($\pm$ 400 Ω·μm). The error range results from using only 4 channels as well as from using large $L$ where $R_{TOT}$ is dominated by the channel resistance.

**3) We etch the MoS$_2$ channel for a constant channel width ($W$).** The uniform MoS$_2$ channels eliminate the possibility of width variation affecting the $R_C$ extraction.

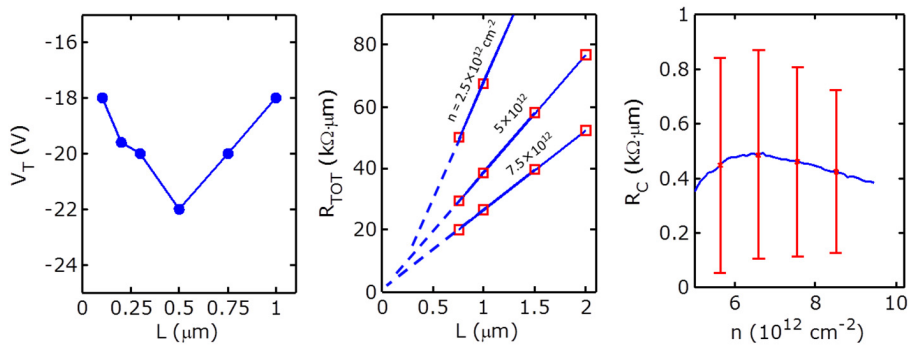

**Fig. S6.** (a) Threshold voltage ($V_T$) vs. $L$ (μm) for a TLM structure on MoS$_2$ using Au contacts deposited at ultra-high vacuum ($P_D = 10^{-9}$ Torr). (b) Total device resistance ($R_{TOT}$) vs channel length ($L$) for a TLM structure on MoS$_2$ using only 4 relatively long channel lengths. (c) Contact resistance ($R_C$) vs carrier density ($n$), as extracted from the TLM in (b), showing the large resulting uncertainty.

## 6. Schottky Barrier Height Extractions

Figure S7 shows the extraction of $\phi_{eff}$ for Au (a-c) and Ni contacts (d–f). The extraction of $\phi_{eff}$ is taken from four $I_D$-$V_G$ curves measured in the range $T = 100$–$200$ K. Extraction of $\phi_{eff}$ using the Richardson plots leads to the plots of $\phi_{eff}$ vs. $E_N$, where $E_N$ is the vertical electric field from the gate electrode to the channel. Note here that $\phi_{eff}$ is an effective barrier height that includes both thermionic and field emission. At the flat-band transition the true Schottky barrier height (due only to thermionic emission) can be extracted, as shown in Figure S8. Since the measured $I_D$ of our devices is limited by gate leakage for $V_G \ll V_T$, we cannot extract $\Phi_B$ for a larger range of $E_N$ that reaches the flat-band voltage, preventing an accurate estimate of $\phi_{eff}$. Thus we take $\Phi_B = 150$ meV for Ni and Au contacts, which has been measured elsewhere.[1,2] An in-depth discussion of $\Phi_B$ extractions from Schottky barrier FETs can be found elsewhere.[3]

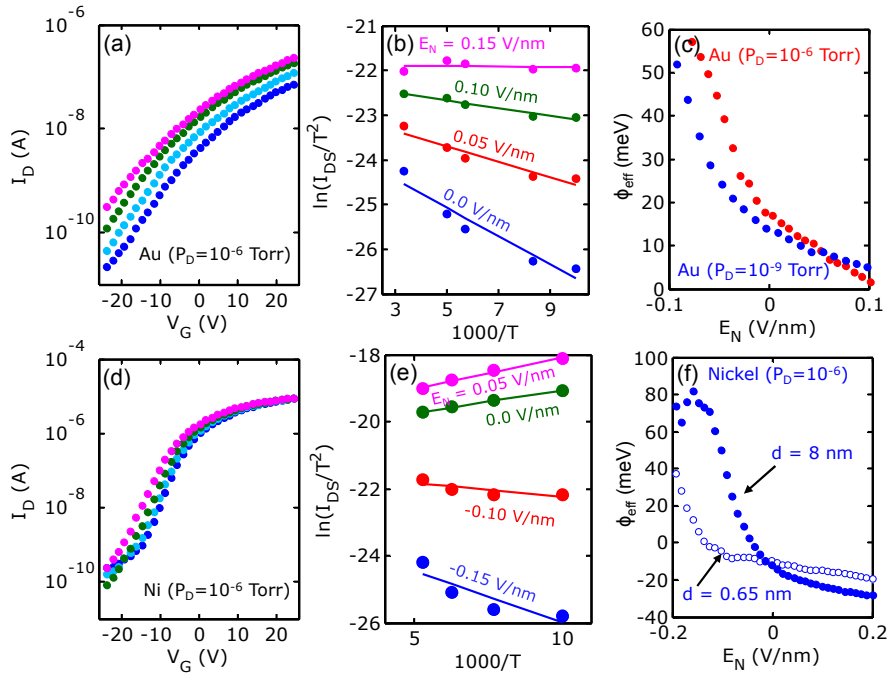

**Fig. S7:** (a) Transfer curve ($I_D$ vs. $V_G$) at $V_{DS} = 100$ mV for Au-contacted MoS$_2$ ($P_D = 10^{-6}$ Torr) measured from $T = 100$-$200$ K. (b) Richardson plot for Au contacts ($P_D = 10^{-6}$ Torr). (c) $\phi_{eff}$ vs. $E_N$ for Au ($P_D = 10^{-6}$ Torr, red) and Au ($P_D = 10^{-9}$ Torr, blue). (d) Transfer curve ($I_D$ vs. $V_G$) at $V_{DS} = 100$ mV for Ni-contacted MoS$_2$ ($P_D = 10^{-6}$ Torr) measured from $T = 100$-$200$ K. (e) Richardson plot for Ni contacts. (f) $\phi_{eff}$ vs. $E_N$ for Ni contacts ($P_D = 10^{-6}$ Torr) for a monolayer and multilayer device.

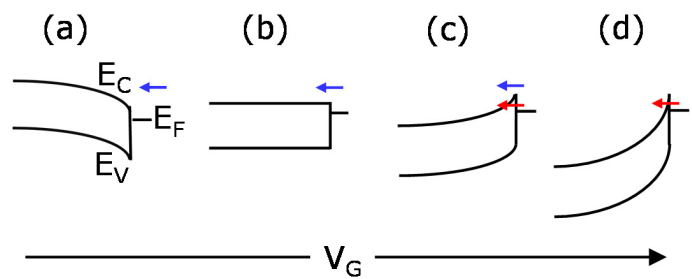

**Fig. S8:** Band diagrams of the MoS$_2$ channel at different stages of the $\Phi_B$ extraction. Blue (red) arrows represent thermionic (field) emission. (a) Only thermionic emission contributes, and changes in $V_G$ (or $E_N$) produce large changes in $\phi_{eff}$. (b) $\phi_{eff}$ equals the true Schottky barrier height ($\Phi_B$) at the flat-band transition. (c) Field emission begins to contribute, resulting in a gradual decrease in $\phi_{eff}$ with increasing $E_N$ as shown here and in Figures S7c,f. (d) Field-emission dominates.

## 7. Metal-MoS$_2$ Interface Resistance Model

This model assumes a parabolic energy dispersion in the contact metal and MoS$_2$, described by three effective masses: one for the 3D density of states, one for the 2D density of the modes in the parallel plane, and one for the tunneling probability. Conservation of the parallel wavevector, $k_\parallel$ defined in the plane of the metal-MoS$_2$ interface, allows integration over the transverse wavevector, $k_\perp$ perpendicular to the metal-MoS$_2$ interface. Thus the current density ($J$) is:

$$J = \frac{qm^*kT}{2\pi^2\hbar^3}\int d\varepsilon_\perp T(\varepsilon_\perp)\log\left(\frac{1-\exp\left(\frac{E_{F1}-\varepsilon_\perp}{kT}\right)}{1+\exp\left(\frac{E_{F2}-\varepsilon_\perp}{kT}\right)}\right)$$

Where $T(\varepsilon_\perp)$ is the transmission probability at transverse energy $\varepsilon_\perp = \hbar^2 k_\perp^2/2m^*$, $m^*$ is the mode counting effective mass, and $E_{F1}$, $E_{F2}$ are the Fermi energies on each side of the barrier. Note that both thermionic emission and tunneling are accounted for here, allowing for calculations of $R_C$ as a function $T$ and $n$ over wide ranges. Once the potential profile is determined by the 1D Poisson equation using an effective metal work function to match the experimental Schottky barrier height, $T(\varepsilon_\perp)$ can be calculated using the transfer matrix method. In this calculation, the degree of Fermi level pinning is accounted for by shifting the vacuum level and semiconductor surface potential. The pinning results in a small Schottky barrier for electrons, taken to be $\Phi_B$ = 150 meV for Ni and Au contacts.[1,2] The specific interfacial resistivity is $\rho_i = \partial V/\partial J|_{V=0}$, which describes solely interfacial resistance (between the top MoS$_2$ layer and the metal) and can therefore be smaller than the contact resistivity ($\rho_C$) which includes contributions from regions immediately above and below the interface.[4,5] Thus while $\rho_C \geq \rho_i$ (see Figure S10) we use $\rho_i$ to calculate $R_C$ vs. $T$ for simplicity since the determination of $\rho_C$ requires prior knowledge of the inter-layer (between layers) transport underneath the contact.

If the contribution of interlayer transport is small (most current flows in the top layer), then $\rho_i \approx \rho_C$. This is a reasonable assumption here, as our extracted $\rho_i$ and $\rho_C$ only differ by a factor of ~4 over the range of $n$ (i.e. a factor of two in the $R_C$ calculation). Note that there is undetermined uncertainty in $\rho_i$ itself, since $\rho_i$ varies exponentially with $\Phi_B$, which we have assumed to be 150 meV. Regardless of the difference between $\rho_i$ and $\rho_C$, the qualitative trends support our conclusion that access resistance improves with lower contact deposition pressure.

## 8. Contact Morphology

The most common adsorbate from air is water, and has been shown to be the primary cause of hysteresis in MoS$_2$ FETs.[6] In addition, the removal of water adsorbates by heating can result in more than 10 times improvement in the mobility as a result of decreased Coulomb impurity scattering in the MoS$_2$ channel.[6-8] However, adsorbates at the metal-MoS$_2$ interface cannot be removed by heating, and thus permanently increase scattering underneath the contacts. MoS$_2$ is particularly hydrophilic owing to the polarity of the sulfur surface[9] indicating that the amount of adsorbed water will depend on the environmental pressure if the MoS$_2$ is not capped with a dielectric. In our study, an ultra-low deposition pressure reduces the amount of adsorbates (i.e. water, oxygen) at the metal-MoS$_2$ interface, leading to an increase in $\mu_C$. This improved transport decreases the access resistance into the channel, lowering the $R_C$.

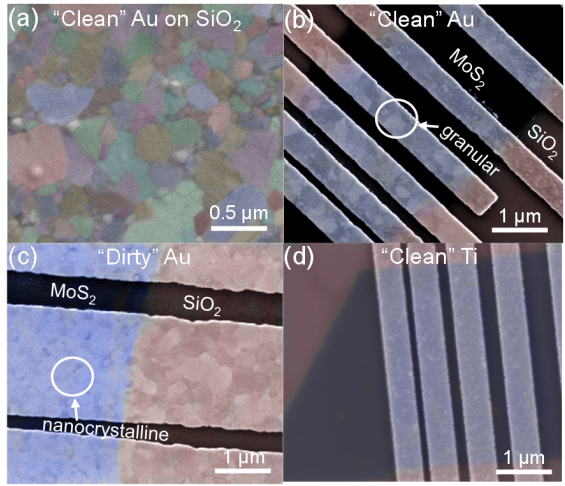

**Fig. S9:** (a) Colorized scanning electron microscope (SEM) image of the morphology of Au electrodes deposited at high vacuum ($P_D = 10^{-9}$ Torr). The Au film is 40 nm thick. (b) Image of Au electrodes ($P_D = 10^{-9}$ Torr) on $MoS_2$ and $SiO_2$. The Au grain size appears large on the $MoS_2$. (c) Image of Au electrodes ($P_D = 10^{-6}$ Torr) on $MoS_2$ and $SiO_2$. The Au grains appear nanocrystalline on the $MoS_2$. (d) Colorized SEM image of Ti contacts ($P_D = 10^{-9}$ Torr) on $MoS_2$ and $SiO_2$ (20 nm Ti with 30 nm Au capping layer).

The different morphologies of Au contacts deposited at high and low $P_D$ offer further insight into the effect of interfacial adsorbates on $R_C$. Au contacts deposited on $SiO_2$ typically show a granular structure regardless of $P_D$ (Figure S9a). When deposited under clean conditions ($P_D = 10^{-9}$ Torr), this granular structure is maintained on the $MoS_2$ (Figure S9b). However, under less clean conditions ($P_D = 10^{-6}$ Torr), the granularity is more nanocrystalline (Figure S9c), which has been observed elsewhere.[10] Typically, Ti, Ni, and Sc contacts also lack any observable granularity (Figure S9d). The improved Au morphology at lower deposition pressure could result from fewer interface adsorbates causing grain nucleation. More specifically, lower deposition pressures appear to lower the Au-$MoS_2$ binding energy ($E_B$) relative to the Au cohesive energy ($E_C$), resulting in larger clustering.[10] We believe that the larger grain sizes result in a smoother interface, increasing $\mu_C$ and lowering the access resistance to the channel.

While this study has focused on device demonstrations, understanding the contact interface from a surface science perspective will require further investigation. This could be achieved, for example using in situ X-ray photoelectron spectroscopy (XPS) during contact deposition[11], residual gas analysis[12], or other surface analysis techniques for examining the formation and composition of adsorbates on the surface of $MoS_2$.

## 9. Specific Interfacial Resistivity ($\rho_i$) vs. Contact Resistivity ($\rho_C$)

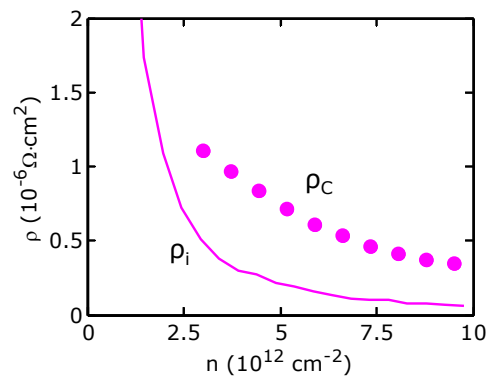

**Fig. S10:** Specific interfacial resistivity ($\rho_i$, line) and contact resistivity ($\rho_C$, solid circles) vs. carrier density ($n$) for clean Au contacts to $MoS_2$.

As discussed in the manuscript, the specific *interfacial* resistivity ($\rho_i$) refers solely to transport through the Schottky barrier at the metal-$MoS_2$ interface. However, the specific contact resistivity ($\rho_C$) accounts for both the Schottky barrier and any vertical, interlayer transport under the contact. Thus, in general, $\rho_C > \rho_i$. Measurements of $\rho_C$ and calculations of $\rho_i$ for clean Au contacts to $MoS_2$ are shown in Figure S10. Clear-ly, $\rho_C > \rho_i$, indicating a contribution from the interlayer resistance. However, the similarity is remarkable considering the exponential dependence of $\rho_i$ on the choice of Schottky barrier height and doping.

## 10. Contact Pitch (CP)

The contact pitch (CP), an important industry metric for device scaling,[13-15] is roughly equal to $L + L_C$, as shown in Figure S11 for a SOI device. Here, $L$ includes the gate length as well as the spacer regions between the gate and source-drain electrodes, which typically play a role in reducing parasitic capacitances. As our devices are back-gated, our minimum CP for $MoS_2$ FETs (~70 nm) does not include contributions from spacer regions, and these would have to be analyzed (and optimized) separately for a top-gated geometry. Nevertheless, the ~70 nm CP (with 40 nm channel and 20-30 nm contacts) shown in Figure 5 of the main write-up represents the smallest TMD FETs with the shortest contacts studied to date.

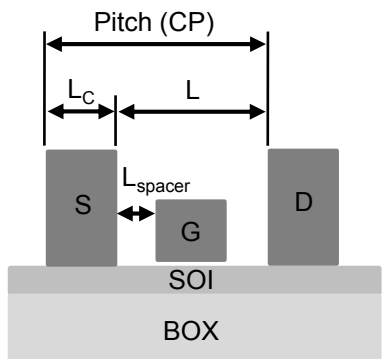

**Fig. S11:** Contact pitch (CP) of a typical FET, here shown for a Si-on-insulator (SOI) device. BOX denotes the buried oxide. The CP is the key parameter determining device density.

## 11. $R_C$ estimates for monolayer $MoS_2$

We were unable to perform accurate TLM $R_C$ extraction on monolayer $MoS_2$ due to device-to-device variation of measured resistances. However, we can estimate an upper bound of the monolayer $R_C$ by subtracting the typical channel resistance of multilayer devices ($R_{CH} = R_{SH}L/W$, based on $R_{SH}$ ~ 30 kΩ/sq. at ~$6 \times 10^{12}$ cm$^{-2}$ carrier density, similar to that of the monolayer devices) from the total device resistance ($R_{TOT}$). The results, shown in Table S1, indicate a large range of $R_C$ values. Since $R_{SH}$ is typically higher ($\mu$ is lower) for monolayer devices,[16] the values in Table S1 and Figure 6a represent upper bounds for the estimates of $R_C$ to monolayer $MoS_2$.

| $L$ (nm) | $R_{TOT}$ (kΩ·μm) | $R_{TOT}/2$ (kΩ·μm) | $R_{SH}L$ (kΩ·μm) | $R_C \approx (R_{TOT}-R_{SH}L)/2$ (kΩ·μm) |
|---|---|---|---|---|
| 200 | 134 | 67 | 6 | 64 |
| 300 | 228 | 114 | 9 | 109.5 |
| 750 | 28 | 14 | 22.5 | 2.75 |
| 1000 | 94 | 47 | 30 | 32 |

**Table S1:** Resistance measurements and extractions for various monolayer $MoS_2$ devices on the same exfoliated flake, with Ni contacts. All resistances are normalized by the device width (here $W \approx 3$ μm).

**Supplementary References**